# Self-supervised Electroencephalogram Representation Learning for Automatic Sleep Staging

**Chaoqi Yang[1], Danica Xiao[2], M. Brandon Westover[3,4], Jimeng Sun[1]**

[1]University of Illinois Urbana-Champaign, IL, United States
[2]Relativity, IL, United States
[3]Harvard Medical School, MA, United States
[4]Beth Israel Deaconess Medical Center, MA, United States

**Corresponding Author:**
Jimeng Sun, PhD
Department of Computer Science and Carle's Illinois College of Medicine
University of Illinois, Urbana-Champaign
201 North Goodwin Avenue Urbana,
Illinois, 61801, United States
Email: jimeng@illinois.edu

## Abstract

**Background:** Deep learning models have shown great success in automating tasks in sleep medicine by learning from carefully annotated Electroencephalogram (EEG) data. However, effectively utilizing a large amount of raw EEG remains a challenge.
**Objective:** In this paper, we aim to learn robust vector representations from massive unlabeled EEG signals, such that the learned vectorized features (1) are expressive enough to replace the raw signals in the sleep staging task; and (2) provide better predictive performance than supervised models in scenarios of fewer labels and noisy samples.
**Methods:** We propose a self-supervised model, named Contrast with the World Representation (ContraWR), for EEG signal representation learning, which uses global statistics from the dataset to distinguish signals associated with different sleep stages. The ContraWR model is evaluated on three real-world EEG datasets that include both at-home and in-lab EEG recording settings.
**Results:** ContraWR outperforms 4 recent self-supervised learning methods on the sleep staging task across 3 large EEG datasets. ContraWR also beats supervised learning when fewer training labels are available (e.g., 4% accuracy improvement when less than 2% data is labeled). Moreover, the model provides informative representative feature structures in 2D projection.
**Conclusions:** We show that ContraWR is robust to noise and can provide high-quality EEG representations for downstream prediction tasks. The proposed model can be generalized to other unsupervised physiological signal learning tasks. Future directions include exploring task-specific data augmentations and combining self-supervised with supervised methods, building upon the initial success of self-supervised learning in this paper.

# Introduction

Deep learning models have shown great success in automating tasks in sleep medicine by learning from high-quality labeled EEG data [1]. EEG data are collected from patients wearing clinical sensors, which generate real-time multi-modal signal data. A common challenge in classifying physiological signals, including EEG, is the lack of enough high-quality labels. This paper introduces a novel self-supervised model that leverages the inherent structure within large, unlabeled, and noisy datasets and produces robust feature representations. These representations can significantly enhance the performance of downstream classification tasks, such as sleep staging, especially in cases where only limited labeled data is available.

Self-supervised learning (specifically, self-supervised contrastive learning) aims at learning a feature encoder that maps input signals into a vector representation using unlabeled data. Self-supervised methods involve two steps: (I) **a pretrain step**: to learn the feature encoder without labels; (II) **a supervised step**: to evaluate the learned encoder with a small amount of labeled data. During the pretext task, some recent methods (e.g., MoCo [2], SimCLR [3]) use the feature encoder to construct positive and negative pairs from the unlabeled data and then optimize the encoder by pushing positive pairs closer and negative pairs farther away. A positive pair consists of two different augmented versions of the same sample (i.e., applying two data augmentation methods separately to the same sample), while a negative pair is generated from two different samples. For example, the augmentation method for EEG data can be denoising or channel flipping. In this practice, existing negative sampling strategies often incur sampling bias [4, 5], especially for noisy EEG data, which significantly hurts performance [6].

Technically, this paper contributes to the pretrain step, where we address the inherent limitations of negative sampling strategy in the existing self-supervised methods (e.g., MoCo [2], SimCLR [3]) by leveraging global data statistics. In contrastive learning, positive pairs bring similarity information, while negative pairs provide contrastive information. Both information are essential in learning an effective feature encoder. This paper proposes a new method to address the limitation in negative sampling, named contrast with the world representation (in abbreviation, ContraWR), where a global average representation over the dataset (called the world representation) is presented as the contrastive information. Therefore, we propose robust contrastive guidance under the absence of labels: *the representation similarity between positive pairs is stronger than the similarity to the world representation*. Derived from global data statistics, the world representation brings robust contrastive information even in noisy environments. Moreover, we strengthen our model with an instance-aware world representation for individual samples, where closer samples have larger weights in calculating the global average. Our experiments show that the instance-aware world representation makes the model more accurate, and this conclusion aligns with the findings from a previous paper [6].

We evaluate the proposed ContraWR on the sleep staging task with three real-world EEG datasets. Our model achieves results comparable to or better than recent popular self-supervised methods, MoCo [2], SimCLR [3], BYOL [7] and SimSiam [8]. The results also show that self-supervised contrastive methods, especially our ContraWR method, are much more powerful in low-label scenarios than supervised learning (e.g., 4% accuracy improvement on sleep staging with less than 2% training data of Sleep EDF dataset).

## Methods

### EEG Datasets

We consider three real-world EEG datasets for this study:
- Sleep Heart Health Study (SHHS) [9, 10] is a multi-center cohort study from the National Heart Lung & Blood Institute assembled to study sleep-disordered breathing, which contains 5,445 recordings. Each recording has 14 Polysomnography (PSG) channels, and the recording frequency is 125.0 Hz. We use the C3/A2 and C4/A1 EEG channels.
- Sleep EDF [11] cassette portion is another benchmark dataset collected in a 1987-1991 study of age effects on sleep in healthy Caucasians aged 25-101 who were taking non sleep-related medications, which contains 153 full-night EEG recordings with recording frequency 100.0 Hz. We extract the Fpz-Cz/Pz-Oz EEG channels as the raw inputs to the model. The first two datasets are all at-home PSG recordings.
- MGH Sleep [1] is collected from sleep laboratory at Massachusetts General Hospital (MGH), where six EEG channels (i.e., F3-M2, F4-M1, C3-M2, C4-M1, O1-M2, O2-M1) are used for sleep staging recorded at 200.0 Hz frequency. After filtering out mismatched signals and missing labels, we finally get 6,478 recordings. Dataset statistics can be found in Table 1, and class label distribution is in Table 2.

**Table 1**. Dataset Statistics

| Name | Location | #channels | #recordings | #epochs | Storage |
|---|---|---|---|---|---|
| SHHS | At home | 2 | 5,445 | 4,535,949 | 260 GB |
| Sleep EDF | At home | 2 | 153 | 415,089 | 20 GB |
| MGH Sleep | In lab | 6 | 6,478 | 4,863,523 | 1,322 GB |

Table 2. Class Label Distribution. Format: #epochs (percentage)

| Name | W | N1 | N2 | N3 | R |
|---|---|---|---|---|---|
| SHHS | 1,306,742 (28.8%) | 169,021 (3.7%) | 1,856,130 (40.9%) | 571,191 (12.6%) | 632,865 (14.0%) |
| Sleep EDF | 285,561 (68.8%) | 21,522 (5.2%) | 69,132 (16.6%) | 13,039 (3.2%) | 25,835 (6.2%) |
| MGH Sleep | 2,154,540 (44.3%) | 481,488 (9.9%) | 700,347 (14.4%) | 855,980 (17.6%) | 671,168 (13.8%) |

## Problem Formulation

To set up the experiments, the raw subject EEG recordings are multi-channel brain waves. First, the unlabeled subject recordings are grouped as the **pretrain set**, the labeled recordings are grouped into the **training or test sets**. The training and test sets are usually small, but their EEG recordings are labeled, while the pretrain set contains a large number of unlabeled recordings. Within each set, the long recordings are segmented into disjoint 30-second windows. Each window is called *an epoch*, denoted as $x \in \mathbb{R}^{C \times N}$. Each epoch has the same format: $C$ input channels and $N$ timestamps from each channel.

For these datasets, the ground truth labels were released by the original data publishers. To align with the problem setting, subjects are randomly assigned to pretrain set, training set and test set with different proportions (90%: 5%: 5% for Sleep EDF and MGH, 98%: 1%: 1% for SHHS, since they have different amount of data). All epochs segmented from one subject are placed within the same set. The pretrain set is used for self-supervised learning, so we remove their labels.

In the **pretrain step**, the EEG self-supervised representation learning problem requires building a feature encoder $f(\cdot)$ from the pretrain group (without labels), which maps one epoch $x$ into a vector representation $h \in \mathbb{R}^d$, where $d$ is the feature dimensionality, such that the representation $h$ can replace raw signal for downstream classification tasks. The evaluation of the encoder $f(\cdot)$ is conducted on the training and test data (with labels). We focus on sleep staging as the **supervised step**, where the feature vector of a sample $x$ will be mapped into five sleep cycle labels, awake (W), rapid eye movement REM (R), Non-REM 1 (N1), Non-REM 2 (N2), Non-REM 3 (N3), based on American Academy of Sleep Medicine (AASM) scoring standards [12]. Specifically, based on the feature encoder from the pretrain step, the training set is used to learn a linear model on top of the feature vectors, and the test group is used to evaluate the linear classification performance.

## Background and Concepts

Self-supervised learning happens in the pretrain task, and it uses representation similarity to exploit the unlabeled signals, with an encoder network $f(\cdot): \mathbb{R}^{C \times N} \to \mathbb{R}^d$ and a nonlinear projection network $g(\cdot): \mathbb{R}^d \to \mathbb{R}^m$. For a given signal $x$ from the pretrain dataset, commonly, one applies data augmentation methods $a(\cdot)$ to produce two different modified signals $\tilde{x}_i, \tilde{x}_j$ (after this procedure, the format does not change), which are then transformed into $h_i, h_j \in \mathbb{R}^d$ by $f(\cdot)$ and further into $z_i, z_j \in \mathbb{R}^m$ by $g(\cdot)$. The vectors $z_i, z_j$ are finally normalized with the *L2* norm onto the unit hypersphere $\frac{z}{||z||} \in \mathbb{S}^{m-1}$.

We call $\frac{z_i}{||z_i||}$ the anchor, $\frac{z_j}{||z_j||}$ the positive sample, and these two together are called a positive pair. For a large number of projections $z_k$ derived from other randomly selected signals (by negative sampling strategy), their representation $\frac{z_k}{||z_k||}$ is commonly conceived of as negative samples (though they are random samples), and any one of them together with the anchor is called a negative pair. The loss functions **L** is derived from the similarity comparison between positive pair and negative pairs (e.g., encouraging similarity of positive pairs to be stronger than that of all the negative pairs, referred to as the noise contrastive estimation (NCE) loss [13] in Appendix). A common forward flow of self-supervised learning on EEG signals can be illustrated as,

$$x \Rightarrow_{a(\cdot)} \tilde{x} \Rightarrow_{f(\cdot)} h \Rightarrow_{g(\cdot)} z \Rightarrow_{L2} \frac{z}{||z||} \Rightarrow_{loss} L.$$

For the data augmentation part, this paper uses bandpass filtering, noising, channel flipping, and shifting (see the visual illustrations in supplementary). We conduct ablation studies on the augmentation methods in experiment and provide the implementation details. To reduce clutter, we also use $z$ to denote the *L2* normalized version in the rest of the paper.

### *(I). ContraWR: Contrast with the World Representation*

As mentioned above, negative sampling can introduce bias for the pretrain step and can undermine representation quality. We propose a new self-supervised learning method, Contrast with the World Representation (ContraWR). ContraWR replaces the large number of negative samples with a single average representation over the dataset, called the world representation. The world representation works as a reference to calibrate the model, making the pretrain step more effective and robust to noise. Our loss function in pretrain step follows the principle: the representation similarity between a positive pair should be stronger than the similarity between the anchor and the world representation.

**The world representation.** Assume $z_i$ is the anchor, $z_j$ is the positive sample, and $z_k$ denotes a random sample. We generate an average representation of the dataset, $z_w$ as the only contrastive information. To formalize, we assume $k \sim p(\cdot)$ is the sample distribution over the dataset (i.e., $k$ is the sample index), independent of the anchor $z_i$. The world representation $z_w$ is defined by,

$$z_w = E_{k \sim p(\cdot)}[z_k].$$

Here, we denote $D = [z: ||z|| \leq 1, z \in \mathbb{R}^m]$. Obviously, $z_w \in D$. In the experiment, $z_w$ is approximated by Monte Carlo method within each batch, i.e., we use the average value over the batch $z_w = \frac{1}{M}\sum_{i=1}^{M} z_j$, where $M$ is the batch size.

**Gaussian kernel measure.** We adopt a Gaussian kernel defined on $D$, $sim(x,y): D \times D \rightarrow (0,1]$ as a similarity measure. Formally, given two feature projections $z_i, z_j$ the similarity is defined as,

$$sim(z_i, z_k) = exp\left(-\frac{||z_i - z_k||^2}{2\sigma^2}\right),$$

where $\sigma$ is a hyperparameter.

**Loss function.** For the anchor $z_i$, the positive sample $z_j$ and the world representation $z_w$, we devise a triplet loss,

$$L(i,j) = \left[sim(z_i, z_w) + \delta - sim(z_i, z_j)\right]_+,$$

Where $\delta > 0$ is the empirical margin, a hyperparameter. The loss is minimized over batches, ensuring that the similarity of positive pair $sim(z_i, z_j)$, is larger than the similarity to the world representation $sim(z_i, z_w)$, by a margin of $\delta$.

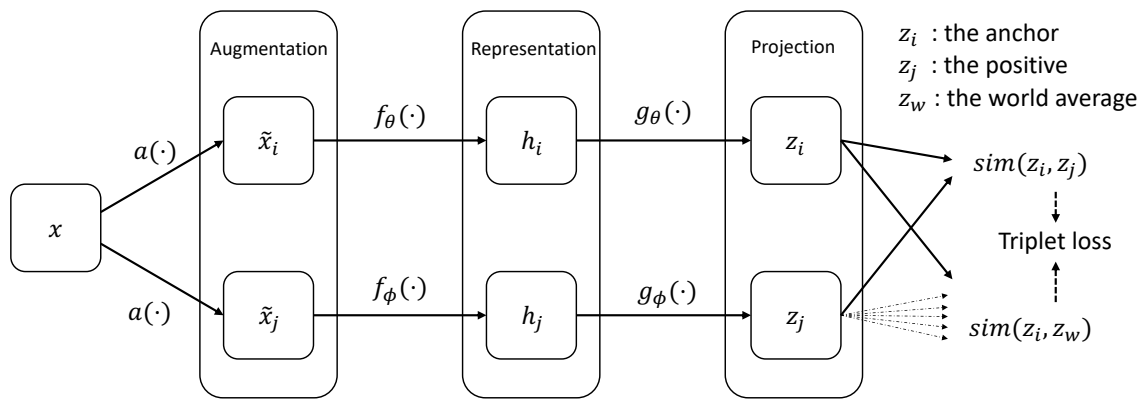

**Figure 1.** ContraWR Model Pipeline. We show the two-way model pipeline in this figure. The online network (upper) is updated by gradient descent, while the target network (lower) is updated by exponential moving average (EMA). Finally, the results from two models form the triplet loss function.

The pipeline of our ContraWR is shown in Figure 1. The online networks $f_\theta(\cdot), g_\theta(\cdot)$ and the target networks $f_\phi(\cdot), g_\phi(\cdot)$ share an identical network structure. Encoder networks $f_\theta(\cdot), f_\phi(\cdot)$ map two augmented versions of the same signal to feature representations, respectively. Then, the projection networks $g_\theta(\cdot), g_\phi(\cdot)$ project the feature representations onto a unit hypersphere, where the loss is defined. During optimization the online networks are updated by gradient descent, and the target networks update parameters from the online network with an exponential moving average (EMA) trick [2].

$$\theta^{(n+1)} \leftarrow \theta^{(n)} - \eta \cdot \nabla_\theta L$$

$$\phi^{(n+1)} \leftarrow \lambda \cdot \phi^{(n)} + (1-\lambda) \cdot \theta^{(n+1)}$$

where $n$ indicates the $n$-th update, $\eta$ is the learning rate, and $\lambda$ is a weight hyperparameter. After this optimization in the pretrain step, the encoder network $f_\theta(\cdot)$ is ready to be evaluated on the training and test sets for the supervised step.

*(II). ContraWR+: Contrast with Instance-aware World Representation*
To learn a better representation, we introduce a weighted averaged world representation, based on the harder principle: the similarity between a positive pair should be stronger than the similarity between the anchor and the weighted average feature representations of the dataset, where the weight is set higher for closer samples. We call the new model ContraWR+. This is a more difficult objective than the simple global average in ContraWR.

**Instance-aware world representation.** In this new model, the world representation is enhanced by modifying the sampling distribution to be instance specific. We define $p(\cdot | z)$ as the instance-aware sampling distribution of an anchor $z$, different from the sample distribution $p(\cdot)$ used in ContraWR,

$$p(\cdot | z) \propto \exp\left(\frac{\langle \cdot, z \rangle}{T}\right),$$

where $T > 0$ is a temperature hyperparameter, such that similar samples are selected with higher probability parametrized by $p(\cdot | z)$. Consequently, for an anchor $z_i$, the instance-aware world representation becomes,

$$z_{w(i)} = E_{k \sim p(\cdot | z_i)}[z_k] = \frac{E_{k \sim p}\left[\exp\left(\frac{\langle z_k, z_i \rangle}{T}\right) \cdot z_k\right]}{E_{k \sim p}\left[\exp\left(\frac{\langle z_k, z_i \rangle}{T}\right)\right]}.$$

Here, $T$ controls the contrastive hardness of the world representation. When $T \to \infty$, $p(\cdot | z)$ is asymptotically identical to $p(\cdot)$, and the above equation reduces to the simple global average form $z_w = E_{k \sim p(\cdot)}[z_k]$; while $T \to 0^+$,, the form becomes trivial, $z_{w(i)} = argmax_{z_k}(sim(z_i, z_k))))$. We have tested different $T$ and find the model is not sensitive to $T$ over a wide range. Here, $z_{w(i)}$ is also approximated by Monte Carlo sampling. We can re-write the similarity measure given the anchor $z_i$ and the new world representation $z_{w(i)}$ as:

$$sim(z_i, z_{w(i)}) = sim\left(z_i, E_{k \sim p(\cdot | z_i)}[z_k]\right)$$

$$= \exp\left(-\frac{1}{2\sigma^2} ||z_i - \frac{E_{k \sim p}\left[\exp\left(\frac{\langle z_k, z_i \rangle}{T}\right) \cdot z_k\right]}{E_{k \sim p}\left[\exp\left(\frac{\langle z_k, z_i \rangle}{T}\right)\right]}||^2\right).$$

In this new method, we also use triplet loss as the final objective.

*Implementations*

**Signal Augmentation.** For the experiments, we use four augmentation methods, illustrated in the supplementary: (I) Bandpass Filtering. To reduce noise, we use an order-1 Butterworth filter (the bandpass is specified in the supplementary); (II) Noising. We add extra high-frequency or low-frequency noise to each channel; (III) Channel Flipping. Corresponding sensors from the left side and the right of the head are swapped; (IV) Shifting. Within one sample, we advance or delay the signal for a certain time span. Detailed configurations of augmentation methods vary for the three datasets, and we list them in the supplementary.

**Baseline Methods.** In the experiments, several recent self-supervised learning methods are implemented for comparison,
- MoCo [2] devises two parallel encoders with exponential moving average (EMA). It also utilizes a large memory table to store new negative samples, which are frequently updated.
- SimCLR [3] uses one encoder network to generate both anchor and positive samples, where negative samples are collected from the same batch.
- BYOL [7] also employs two encoders: one online network and one target network. They put one more predictive layer on top of the online network to predict (reconstruct) the result from the target network, while no negative samples are presented.
- SimSiam [8] uses the same encoder networks on two sides and also does not utilize the negative samples.

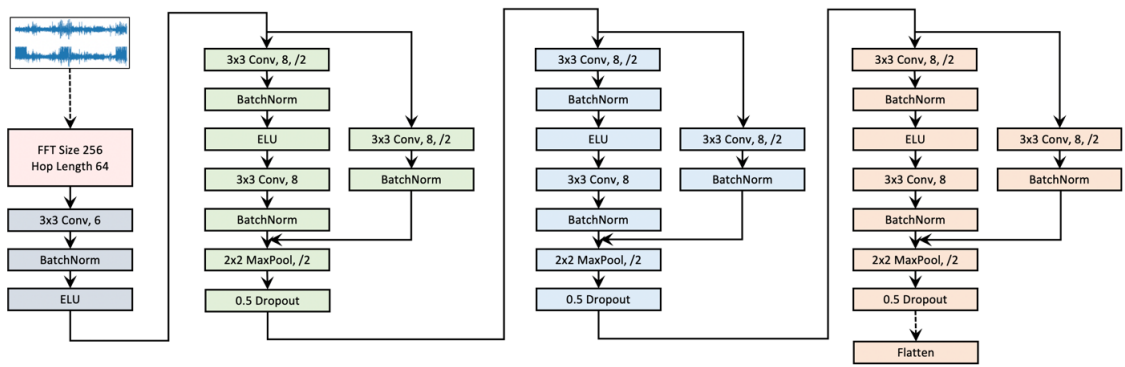

**Figure 2.** STFT Convolutional Encoder Network. The encoder network first transforms raw signals into spectrogram by short time Fourier transform (STFT), then a CNN-based encoder is built on top of the spectrogram.

**Model Architecture.** Our proposed ContraWR and ContraWR+ use the same encoder architecture, as shown in Figure 3. This architecture cascades a short time Fourier transform (STFT) operation, a 2D convolutional neural network layer, and three 2D convolutional blocks. Empirically, we find that apply neural networks on the STFT spectrogram generates better accuracy than on the raw signals. Same practices can be found in [14, 15]. For a fair comparison, the baseline approaches use the same augmentation and encoder architecture.

We also consider a supervised model (called Supervised) with the same encoder, on top of which we further add a 2-layer fully connected network (128-unit for Sleep EDF, 256-unit for SHHS, and 192-unit for MGH) for the sleep staging classification task. The supervised model does not use the pretrain set but is trained from scratch on raw EEG signals in the training set and tested on the test set. We also include an Untrained Encoder model as a baseline, where the encoder is initialized but not optimized in the pretrain step.

**Evaluation Protocol.** We evaluate performance on the sleep staging task with overall five-class classification accuracy. Each experiment is conducted with five different random seeds. For self-supervised methods, we optimize the encoder for 100 epochs (here, "epoch" is a concept in deep learning) with unlabeled data and use the training set to find a good logistic classifier and use the test set data for evaluation following [2, 3]. For the supervised method, we train the model for 100 epochs on the training set. Our setting ensures the convergence of all models.

## Results

### Better Accuracy in Sleep Staging

Comparisons on the downstream sleep staging task are shown in Table 3.

**Table 3.** Sleep Staging Accuracy Comparison with Difference Methods (%). Format: mean ± standard deviation of training/test over 5 random seeds.

| Name | Sleep EDF | SHHS | MGH Sleep |
|---|---|---|---|
| Supervised | 84.98 ± 0.3562 | 75.61 ± 0.9347 | 69.73 ± 0.4324 |
| Untrained Encoder | 77.83 ± 0.0232 | 60.03 ± 0.0448 | 55.64 ± 0.0082 |
| MoCo | 85.58 ± 0.7707 | 77.10 ± 0.2743 | 62.14 ± 0.7099 |
| SimCLR | 83.79 ± 0.3532 | 76.61 ± 0.3007 | 67.32 ± 0.7749 |
| BYOL | 85.61 ± 0.7080 | 76.64 ± 0.3783 | 70.75 ± 0.1461 |
| SimSiam | 84.78 ± 0.8028 | 74.25 ± 0.4796 | 62.08 ± 0.4902 |
| ContraWR | 85.94 ± 0.2326 | 77.52 ± 0.5748 | 71.97 ± 0.1774 |
| ContraWR+ | 86.90 ± 0.2288 | 77.97 ± 0.2693 | 72.03 ± 0.1823 |

All self-supervised methods outperform the Untrained Encoder model, indicating that the pretrain step does learn some useful features from unlabeled data. We observe that ContraWR and ContraWR+ both outperform the supervised model, suggesting that the feature representations provided by the encoder can better preserve the predictive features and filter out noises than using the raw signals for the sleep staging task, in the case when the amount of labeled data available is not sufficient (e.g., less than 2% in Sleep EDF). Compared to other self-supervised methods, our proposed model ContraWR+ also provides better predictive accuracy, i.e., about 1.3% on Sleep EDF, 0.8% on SHHS, 1.3% on MGH Sleep. The performance improvements are mostly significant with p<.001, except the p-values comparing with MoCo on

Sleep EDF dataset is 1.7e-3. MGH Sleep data contains more noise than the other two datasets (reflected by the relatively low accuracy with supervised model on raw signals). It is notable that the performance gain is much more significant on MGH over other self-supervised or supervised models (about 3.3% relative improvement on accuracy) which suggests that the proposed models handle noisy environments better.

*Ablation Study on Data Augmentations*

We also inspect the effectiveness of different augmentation methods on EEG signals, shown in Table 4.

**Table 4.** Evaluation Accuracy of Different Augmentations (%). Format: mean ± standard deviation of training/test over 5 random seeds.

| Augmentations | Accuracy |
|---|---|
| Bandpass | 84.23 ± 0.2431 |
| Noising | 83.60 ± 0.1182 |
| Shifting | 84.65 ± 0.2844 |
| Bandpass + Flipping | 85.77 ± 0.2337 |
| Noising + Flipping | 84.45 ± 0.1420 |
| Shifting + Flipping | 85.13 ± 0.0558 |
| Bandpass + Noising | 85.37 ± 0.1214 |
| Noising + Shifting | 84.78 ± 0.1932 |
| Shifting + Bandpass | 85.25 ± 0.1479 |
| Bandpass + Noising + Flipping | 85.76 ± 0.1794 |
| Noising + Shifting + Flipping | 85.17 ± 0.2301 |
| Shifting + Bandpass + Flipping | 86.38 ± 0.2789 |

We empirically test all possible combinations of four considered augmentations: channel flipping, bandpass filtering, noising, shifting. Since channel flipping cannot be applied solely, we combine it with other augmentations. The evaluation is conducted on Sleep EDF with ContraWR+ model. To sum up, all augmentation methods are beneficial, and collectively, they can further boost the classification performance.

*Varying Amount of Training Data*

To further investigate the benefits of self-supervised learning, we evaluate the effectiveness of the learned feature representations with varying training data on Sleep EDF in Figure 3. The default setting is to split all the data into pretrain/training/test sets by 90%: 5%: 5% (as stated in the problem formulation). In this section, we keep the 5% test set unchanged and re-split the pretrain and training sets (after re-splitting, we ensure the training set data all have labels and remove the labels from the pretrain set), such that the training proportion becomes 0.5%, 1%, 2%, 5%, 10%, and the rest is used for the pretrain set. This "re-splitting" is conducted at the subject level, after which we again segment each subject's recording within the pretrain or training set. We compare our ContraWR+ to MoCo, SimCLR, BYOL, SimSiam, and the supervised baseline models. Our model outperforms the compared models consistently with different amount of training data. For example, our model achieves similar performance (with only 5% data as training) compared to the best baseline, BYOL, which needs twice amount of training data (10% data as training). Also, compared to the supervised model, the self-supervised methods perform better when the labels are insufficient, e.g., only ≤ 2% of the data are labeled.

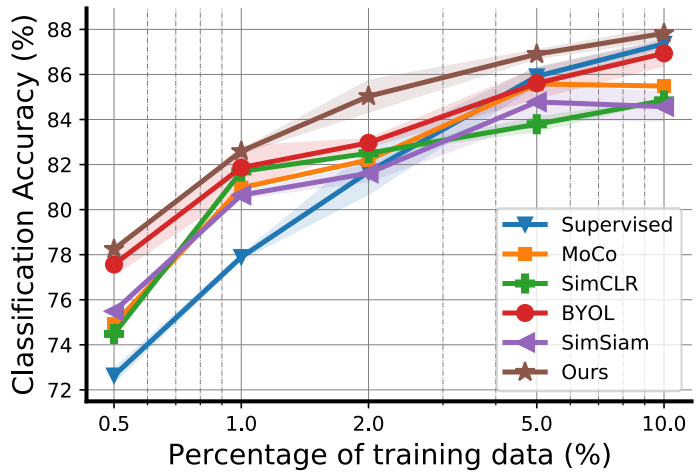

**Figure 3.** Model Performance with Different Amount of Training Data (on Sleep EDF). Format: curves are the mean values and shaded areas are the standard deviation of training/test over 5 random seeds. All models have the same encoder network architecture. For the self-supervised method, we train a logistic regression model on top of the frozen encoder with the training set, and for the supervised model, we train the encoder along with the final nonlinear classification layer from scratch with the training set. The amount of training data is set 0.5%, 1%, 2%, 5%, 10%. Each configuration runs with 5 different random seeds and the error bars indicate the standard deviation over 5 seeds.

*Representation Projection*

We next sought to assess the quality of the learned feature representations. To do this, we use the representations produced by ContraWR+ on the MGH dataset and randomly select 5,000 signal epochs per label from the dataset. The ContraWR+ encoder is optimized on the pretrain step without using the labels. We extract feature representations for each sample through the encoder network and use uniform manifold approximation and projection (UMAP) [16] to project onto the 2D space. We finally color code samples according to sleep stage labels for illustration.

The 2D projection is shown in Figure 4. We also compute the confusion matrix from the evaluation stage (based on the test set), also shown in Figure 4. In the UMAP projection, epochs from the same latent class are closely co-located, which means the pretrain step extracts important information for sleep stage classification from the raw unlabeled EEG signals. Stage N1 overlaps with stages W, N2, and N3, which is as expected given that N1 is often ambiguous and thus difficult to classify even for well-trained experts [1].

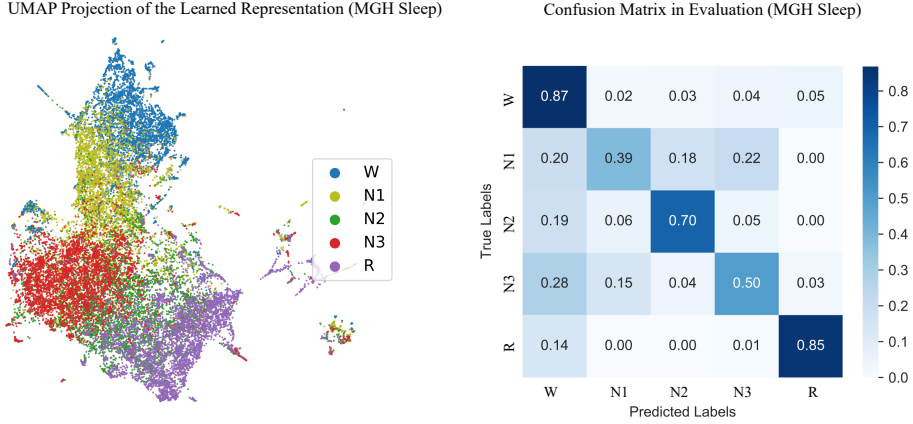

**Figure 4.** UMAP Projection and Confusion Matrix. Using MGH dataset, we project the output representations of each signal into 2D space and color by the actual labels (left). We also show the confusion matrix on sleep staging (right).

*Hyperparameter Ablation Study*
To investigate the sensitivity of our model to hyperparameter settings, we test with different batch sizes and train on different values for the Gaussian parameter $\sigma$, temperature $T$, and margin $\delta$. We focus on the ContraWR+ model and evaluate it on the Sleep EDF dataset. During the experiment, the default settings are batch size = 256, $\sigma = 2, T = 2, \delta = 0.2$, learning rate $\eta$ = 2e-4, weight decay = 1e-4, epoch = 100. When testing on one hyperparameter, others are held fixed.

Ablation study results are in shown in Figure 5; the red star indicates the default configuration. Each configuration runs with 5 different random seeds and the error bars indicate the standard deviation over 5 experiments. We see that the model is not sensitive to batch size. We see that over a large range (< 10) the model is insensitive to the Gaussian width $\sigma$. For temperature $T$, we noted previously that a very small $T$ may be problematic, and a very large $T$ reduces ContraWR+ to ContraWR. Based on the ablation experiments the performance is relatively insensitive to choices of $T$. For the margin $\delta$, the distance difference is bounded (given fixed $\sigma = 2$),

$$||sim(z_i, z_w) - sim(z_i, z_j)|| \leq ||\exp\left(-\frac{0^2}{2\sigma^2}\right) - \exp\left(-\frac{2^2}{2\sigma^2}\right)||^2 \approx 0.3935.$$

Thus, $\delta$ should be chosen large enough, i.e., $\delta \geq 0.1$.

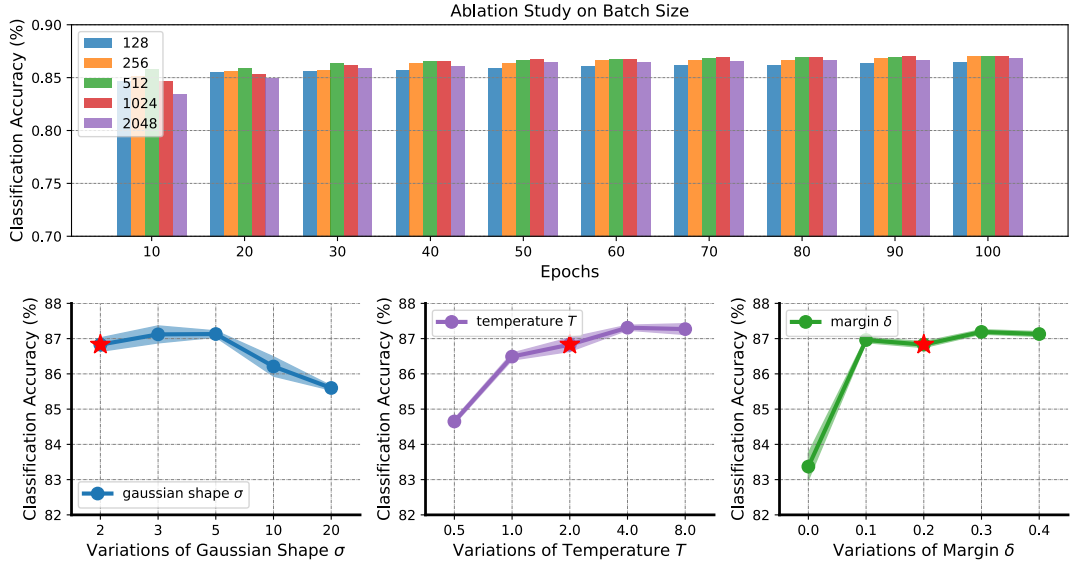

**Figure 5.** Ablation Study on Batch Size and Three Hyperparameters. Format: curves are the mean values and shaded areas are the standard deviation of training/test over 5 random seeds. The red star denotes the default setting. It is obvious that with a larger batch size, the model will perform better, while it is not sensitive to all hyperparameters.

## Discussion

### Principle Results

Our proposed ContraWR and ContraWR+ models outperform 4 recent self-supervised learning methods on the sleep staging task across 3 large EEG datasets (with p<.002). ContraWR+ also beats supervised learning when fewer training labels are available (e.g., 4% accuracy improvement when less than 2% data is labeled). Moreover, the models provide well-separated representative structures in 2D projection.

### Comparison with Prior Work

#### *Self-supervised Learning*

Many deep generative methods have been proposed for unsupervised representation learning. They mostly rely on auto-encoding [17-19] or adversarial training [20-22]. Mutual information (MI) maximization is also popular for compressing input data into a latent representation [23-25].

Recently, self-supervised contrastive learning [2, 3, 7, 8, 14] has become popular, where loss functions are devised from representation similarity and negative sampling. However, one recent work [4] highlighted inherent limitations of negative sampling and showed that this strategy could hurt the learned representation significantly [5]. To address these limitations, Chuang et al. [5] utilized the law of total probability and approximated the per-class negative sample distribution using the weighted sum of the global data distribution and the expected class label distribution.

However, without the actual labels, the true class label distribution is unknown. Grill et al. [7] and Chen et al. [8] proposed ignoring negative samples and learning latent representations using only positive pairs.

In this paper, we still use the negative information by replacing negative sampling with the global average (i.e., the world representation). We argue and provide experiments showing that contrasting with the world representation is more powerful and robust in the noisy EEG setting.

### *EEG Sleep Staging*
Before the emergence of deep learning, several traditional machine learning approaches [26-28] significantly advanced the field using hand-crafted features, as highlighted in [29]. Recently, deep learning models have been applied to various large sleep databases. SLEEPNET [29] built a comprehensive system combining many machine learning models to learn sleep signal representations. Biswal et al. [1] designed a multi-layer RCNN model to process multi-channel signals from EEG. To provide interpretable stage prototypes, Al-Hussaini et al. [30] developed a SLEEPER model that utilizes a particular deep learning approach called prototype learning guided by a decision tree to provide more interpretable results. These works rely on a large set of labeled training data. However, the annotations are expensive, and often times the labeled set is small. In this paper, we exploit the large set of unlabeled data to improve the classification, which is more challenging.

### *Self-supervised Learning on Physiological Signals*
While image [31, 32], video [33], language [34, 35], and speech [36] representations have benefited from contrastive learning, research on learning physiological signals has been limited [37, 38]. Lemkhenter et al. [39] proposed phase and amplitude coupling for physiological data augmentation. Banville et al. [40] conducted representation learning on EEG signals, and they targeted monitoring and pathology screening tasks, without utilizing frequency information. Cheng et al. [41] learned subject-aware representations for ECG data and tested various augmentation methods. While most of these methods are based on pairwise similarity comparison, our model brings contrastive information from global data statistics, providing more robust representations. Also, we extract signal information from the spectral domain.

### Strengths and Limitations
Strengths of our study are: (I) we use three real-world datasets collected from different institutes and across different year ranges, and two are publicly available; (II) our PSG recordings are diverse and generalizable, including two datasets collected at home and one collected in the lab setting, all have relatively large sizes; (III) we have open sourced our data processing pipelines and all programs used for his study, including the baseline model implementations; (IV) we propose new data augmentation methods for PSG signals and have systematically evaluated their effectiveness. However, limitations of our study should be noted, and they include the following: (I) we fixed the neural network encoder architecture in the study, which we plan to explore using other models like recurrent neural networks in the future.;

(II) we have utilized the STFT to extract spectrograms, but we may consider alternative techniques such as wavelet transformation in future; (III) our current data augmentation methods are based on clinical knowledge, and we aim to investigate data-driven approaches to design more effective methods in the future.

## Conclusions

This paper is motivated by the need to learn effective EEG representations from large unlabeled noisy EEG datasets. We propose a self-supervised contrastive method, ContraWR, and its enhanced variant, ContraWR+. Instead of creating a large number of negative samples our method contrasts samples with an average representation of many samples. The model is evaluated on a downstream sleep staging task with three real-world EEG datasets. Extensive experiments show that the model is more powerful and robust than multiple baselines including MoCo, SimCLR, BYOL, and SimSiam. ContraWR+ also outperforms the supervised counterpart in label-insufficient scenarios.

## Acknowledgements

Chaoqi Yang implemented the method and conducted the experiments. All authors were involved in developing the ideas and writing the paper.

This work was in part supported by the National Science Foundation award SCH-2014438, IIS-1418511, CCF-1533768, IIS-1838042, the National Institute of Health (R01NS107291, R56HL138415, 1R01NS102190, 1R01NS102574, RF1AG064312), the Glenn Foundation for Medical Research and the American Federation for Aging Research (Breakthroughs in Gerontology Grant), and the American Academy of Sleep Medicine (AASM Foundation Strategic Research Award).

## Conflicts of Interest

MBW is co-founder of Beacon Biosignals, which played no role in this study. The other authors have no competing interests to declare.

## Abbreviations

EEG: Electroencephalogram
ContraWR: contrast with the world representation
EMA: exponential moving average
SHHS: sleep heart health study
PSG: Polysomnography
MGH: Massachusetts General Hospital
UMAP: uniform manifold approximation and projection
REM: rapid eye movement
AASM: American Academy of Sleep Medicine
NCE: noise contrastive estimation
STFT: short time Fourier transform

## Multimedia Appendix 1

Theoretical loss boundness analysis. https://drive.google.com/file/d/1-2yNzy0y6Q8Zho1loqbzI3xXbEY9U8Z4/view

**Multimedia Appendix 2**

Powerpoint presentation slides. [https://docs.google.com/presentation/d/1CA-bCxpRR5Mets2Nwshjlp6PsrRmN9mD0RQjuCHr1Tg/edit?usp=sharing](https://docs.google.com/presentation/d/1CA-bCxpRR5Mets2Nwshjlp6PsrRmN9mD0RQjuCHr1Tg/edit?usp=sharing)

**Multimedia Appendix 3**

Video presentation link.
[https://drive.google.com/file/d/14BVlVzYSB10vF49QYQmcs93UxioDODAl/view](https://drive.google.com/file/d/14BVlVzYSB10vF49QYQmcs93UxioDODAl/view)

**Multimedia Appendix 4**

Open-source GitHub repository, including data process pipeline for Sleep EDF and SHHS datasets, scripts for ContraWR, ContraWR+, and four self-supervised baseline models. [https://github.com/ycq091044/ContraWR](https://github.com/ycq091044/ContraWR).

**Multimedia Appendix 5**

Illustration for data augmentations (bandpass filtering, noising, flipping, shifting). [https://drive.google.com/file/d/1CaagQm8O2vCzux3Msh7wkoucZhY6pX8V/view?usp=sharing](https://drive.google.com/file/d/1CaagQm8O2vCzux3Msh7wkoucZhY6pX8V/view?usp=sharing)

**Multimedia Appendix 6**

Supplementary on model implementation.
[https://drive.google.com/file/d/1ZBu_Bzfa_EY3RyD1DZGrCNzTj1C7T8ZW/view?usp=sharing](https://drive.google.com/file/d/1ZBu_Bzfa_EY3RyD1DZGrCNzTj1C7T8ZW/view?usp=sharing)